\documentclass[a4paper, 12pt]{article}
\usepackage{amssymb, amsmath, graphicx}
\usepackage[left=10mm, top=10mm, right=10mm, bottom=15mm]{geometry}

\title{Charge in electric field in probability representation}
\author{V. I. Man'ko \and E. V. Shchukin}
\date{}

\begin{document}
\maketitle

\begin{abstract}
Green function and linear integrals of motion for a charged particle moving in electric field
are discussed. Wigner function and tomogram of the ststionary states of the charge are
obtained. Connection of quantum propagators for Schr\"{o}dinger evolution equation, Moyal
evolution equation and evolution equation in tomographic probability representation for charge
moving in electric field is discussed.
\end{abstract}

\section{Introduction}

The quantum mechanics formalism is based on the notions of wave function \cite{adp-79-489,
cr-183-447, cr-184-273, cr-185-380, zfp-40-332} and density matrix
\cite{neumann-mathematische-grundlagen-quantenmechanik}. The possible interpretations of
quantum properties including the interpretation in terms of hidden variables have been
suggested in \cite{pr-85-166, pr-85-180} and analysis of this approach is presented in
\cite{bell-speakable-unspeakable-quantum-mechanics,
terletskii-gusev-problems-causality-quantum-mechanics,
wheeler-zurek-quantum-theory-measurement}. The tomographic map \cite{qso-7-615} gives the
possibility to formulate quantum mechanics using the notion of probability and, in principle,
avoiding the notion of wave function \cite{fp-27-801, pl-a-213-1, jrlr-18-407, jrlr-17-579}.
The attempts to construct quantum formalism similar to classical one done in the spirit of
Moyal approach \cite{pcps-45-99} based on Wigner function \cite{pr-40-749} were made using
quasidistributions of different kind \cite{prl-10-84, prl-10-277, ppmsj-23-264, ap-190-107}.
The evolution equation of such quasidistributions \cite{pr-a-59-971, fp-27-801} provides the
complete information on behaviour of the quantum systems which is equivalent to the
information contained in von Neumann equation for density matrix
\cite{neumann-mathematische-grundlagen-quantenmechanik}. There are several modern quantum
problems \cite{prl-70-1244, pl-a-219-180, prl-76-1796, nc-b-110-545, pl-a-228-29, qso-9-381,
prl-77-2198, pr-a-54-4560, pr-a-53-4528, prl-76-4344, jp-a-34-6185} related to trapped ions,
photons in cavity, quantum computers as well as traditional ones \cite{jpif-6-861, zp-45-430}
which can be considered in framework of the probability representation of quantum mechanics.
In this work we analyse one of these problems for charge moving in electric field.

One of the solvable problems of quantum motion is the problem of charge moving in electric
field. It is important problem and we will discuss this problem in framework of tomographic
probability representation of quantum states \cite{fp-17-397, pr-a-40-2847, prl-70-1244,
qso-7-615, qso-8-1017, pr-a-53-2998, el-37-79, prl-76-4344, jmo-44-2281}.

The probability representation of quantum mechanics was introduced in \cite{fp-27-801,
pl-a-213-1}. The Wigner function \cite{pr-40-749} of quantum states is the informative
characteristic of the state. The optical tomography scheme \cite{prl-70-1244} was used to
reconstruct the Wigner function of photons. Extension of the optical tomography method
containing two real parameters was suggested in \cite{qso-7-615}. The tomographic probability
distribution of quantum state was employed in \cite{fp-27-801, pl-a-213-1} to replace the wave
function and density matrix by the probability density which determines the state. This
approach to description of quantum states created new formalism of quantum mechanics similar
to classical statistics. The evolution \cite{fp-27-801, pl-a-213-1, pr-a-59-1809, jrlr-17-579}
and energy levels \cite{jrlr-18-407, jrlr-17-579} in tomographic representation correspond to
Moyal formulation of quantum mechanics \cite{pcps-45-99}. The probability representation was
introduced for spin in \cite{pl-a-229-335, zetf-112-796, jetp-114-437, pr-a-57-671,
prl-84-802}. The Pauli equation in tomographic representation was obtained in
\cite{jp-a-34-3461}. There are quasidistribution functions which correspond to density matrix
in different representations like Wigner function \cite{pr-40-749}, Glauber-Sudarshan
$P$-function \cite{prl-10-84} and \cite{prl-10-277}, Husimi $Q$-function \cite{ppmsj-23-264}.
The tomograms have specific property of standard probability distribution and they completely
describe the quantum states. Many physical systems are described by the quadratic Hamiltonians
\cite{malkin-manko-dynamical-symmetries-coherent-states-quantum-systems}. The charge moving in
homogeneous electric field is important example of such systems. The Gaussian and
Gauss-Hermite states for such system were studied using time-dependent integrals of motion
linear in position and momentum in \cite{pl-a-30-414}. The wave functions of Gaussian states
and number states can be obtained for the charge moving in electric field in explicit form.
The tomographic description of the states is the main goal of our work. We will obtain
tomograms and propagator for the evolution equation in the probability representation for the
charge moving in electric field. We will also discuss the connection of Green function for
Schr\"{o}dinfer equation and the propagator.

\section{Green function and integrals of motion}

First we discuss the Green function of Schr\"{o}dinger evolution equation for the wave
function in position representation
\begin{equation}\label{e-Schrodinger}
    i\hbar\frac{\partial\psi(x,t)}{\partial t} = \widehat{H}(t)\psi(x,t).
\end{equation}
Unitary operator $\widehat{U}(t)$ is called evolution operator of this equation if for any
solution $\psi(x,t)$ expressed in terms of the state vector $|\psi,t\rangle$ one has the
relation
\begin{equation}\label{e-U}
    |\psi,t\rangle = \widehat{U}(t)|\psi,0\rangle.
\end{equation}
From this equation follows that at the time moment $t=0$ the evolution operator coincides with
the identity operator:
\begin{equation}
    \widehat{U}(0) = \widehat{1}.
\end{equation}
The matrix element of the operator $\widehat{U}(t)$ in position representation is called
"Green function" of the Schr\"{o}dinger evolution equation \eqref{e-Schrodinger}
\begin{equation}
    G(x,x^{\prime},t) = \langle x |\widehat{U}(t)| x^{\prime} \rangle.
\end{equation}
In position representation the equation \eqref{e-U} can be rewritten in the integral form
\begin{equation}\label{e-Gpsi}
    \psi(x,t) = \int G(x,x^{\prime},t)\psi(x^{\prime},0)\,dx^{\prime}.
\end{equation}
This equality means that the Green function is the kernel of the evolution operator. It is
obvious that for the initial time moment the Green function coincides with Dirac
delta-function, i.e.,
\begin{equation}\label{e-Gi}
    G(x,x^{\prime},0) = \delta(x-x^{\prime}).
\end{equation}
It is easy to obtain the evolution equation for the Green function $G(x,x^{\prime},t)$.
Substituting expression \eqref{e-Gpsi} into Schr\"{o}dinger equation for the wave function we
get
\begin{equation}
    i\hbar\frac{\partial\psi}{\partial t}(x,t) =
    i\hbar\int\frac{\partial G}{\partial t}(x,x^{\prime},t)\psi(x^{\prime},0)\,dx^{\prime} =
    {\widehat{H}(t)}_{[x]}\psi(x,t) =
    \int{\widehat{H}(t)}_{[x]}G(x,x^{\prime},t)\psi(x^{\prime},0)\,dx^{\prime},
\end{equation}
where notation $\widehat{H}_{[x]}$ means that the energy-operator $\widehat{H}$ acts on the
variable $x$ of the Green function. Since the above equations are valid for arbitrary wave
function we obtain the following equation for the Green function
\begin{equation}\label{e-Gt}
    i\hbar\frac{\partial G}{\partial t}(x,x^{\prime},t) =
    {\widehat{H}(t)}_{[x]}G(x,x^{\prime},t).
\end{equation}

Now we discuss the notion of quantum integrals of motion. The operator $\widehat{I}(t)$ is
called the integral of motion if its mean does not depend on time, i.e.
\begin{equation}
    {\left\langle\widehat{I}(t)\right\rangle}_{\psi} =
    \langle\psi,t|\widehat{I}(t)|\psi,t\rangle
\end{equation}
for any solution $|\psi,t\rangle$ of the Schr\"{o}dinger equation. It means that
\begin{equation}
    \frac{d}{dt}{\left\langle\widehat{I}(t)\right\rangle}_{\psi} =
    \frac{d}{dt}\langle\psi,t|\widehat{I}(t)|\psi,t\rangle = 0.
\end{equation}
Since the total time derivative can be expressed in the form containing the term with the
commutator
\begin{equation}
    \frac{d}{dt}{\left\langle\widehat{I}(t)\right\rangle}_{\psi} =
    {\left\langle\frac{\partial\widehat{I}(t)}{\partial t}+
    \frac{i}{\hbar}[\widehat{H},\widehat{I}(t)]\right\rangle}_{\psi},
\end{equation}
the operator $\widehat{I}(t)$ is integral of motion iff it satisfies the evolution equation of
the form
\begin{equation}\label{e-I}
    i\hbar\frac{\partial\widehat{I}(t)}{\partial t} = [\widehat{H},\widehat{I}(t)],
\end{equation}
in other words, the operator $\widehat{I}(t)$ is the integral of motion iff its total time
derivative is equal to zero:
\begin{equation}
    \frac{d\widehat{I}(t)}{dt} = \frac{\partial\widehat{I}(t)}{\partial t} +
    \frac{i}{\hbar}[\widehat{H},\widehat{I}(t)] = 0.
\end{equation}
By definition of the evolution operator $\widehat{U}(t)$ one has the equalities
\begin{equation}
    |\psi,t\rangle = \widehat{U}(t)|\psi,0\rangle, \qquad
    \langle\psi,t| = \langle\psi,0|\widehat{U}^+(t) = \langle\psi,0|\widehat{U}^{-1}(t).
\end{equation}
We used the property of unitarity of the evolution operator which takes place for Hermitian
Hamiltonians. By definition of the integral of motion $\widehat{I}(t)$ we have also the
equalities
\begin{equation}
    \langle\psi,0|\widehat{I}(0)|\psi,0\rangle = \langle\psi,t|\widehat{I}(t)|\psi,t\rangle =
    \langle\psi,0|\widehat{U}^{-1}(t)\widehat{I}(t)\widehat{U}(t)|\psi,0\rangle,
\end{equation}
from which one gets
\begin{equation}
    \widehat{I}(0) = \widehat{U}^{-1}(t)\widehat{I}(t)\widehat{U}(t)\quad\text{or}\quad
    \widehat{I}(t) = \widehat{U}(t)\widehat{I}(0)\widehat{U}^{-1}(t).
\end{equation}
This mean that the evolution operator and the integral of motion satisfy the operator equation
\begin{equation}
    \widehat{I}(t)\widehat{U}(t) = \widehat{U}(t)\widehat{I}(0).
\end{equation}
The matrix elements of the equal operators must be equal, i.e.,
\begin{equation}\label{e-IU}
    {\left(\widehat{I}(t)\widehat{U}(t)\right)}_{xx^{\prime}} =
    {\left(\widehat{U}(t)\widehat{I}(0)\right)}_{xx^{\prime}}.
\end{equation}
The above equation is written in position representation. Note that for any operator
$\widehat{f}$ and function $\psi(x)$ the following equality takes place
\begin{equation}
    (\widehat{f}\psi)(x) = \int f(x,x^{\prime})\psi(x^{\prime})\,dx^{\prime},
\end{equation}
which is an analog of finite-dimensional equality
\begin{equation}
    {f(l)}_i = \sum\limits^n_{j=1}f_{ij}l_j,
\end{equation}
where $f: L \to L$ is linear operator in n-dimensional linear space $L$, $l \in L$ is
$n$-vector, $f_{ij}$ and $l_i$ are matrix elements of operator $f$ and components of vector
$l$, respectively, with respect to some fixed basis of linear space $L$. Equation \eqref{e-IU}
can be presented in the form
\begin{equation}
\begin{split}
    \int I(x,y,t)G(y,x^{\prime},t)\,dy &= {\widehat{I}(t)}_{[x]}G(x,x^{\prime},t) =
    \int G(x,y,t)I(y,x^{\prime},0)\,dy = \int I(y,x^{\prime},0)G(x,y,t)\,dy \\
    &= \int I^t(y,x^{\prime},0)G(x,y,t)\,dy =
    {\widehat{I}^t(0)}_{[x^{\prime}]}G(x,x^{\prime},t).
\end{split}
\end{equation}
Here $\widehat{I}^t$ implies the transposed operator. Thus, we obtain system of two equations
for Green function in case of physical system with one degree of freedom
\begin{equation}
    {\widehat{I}(t)}_{[x]}G(x,x^{\prime},t) =
    {\widehat{I}^t(0)}_{[x^{\prime}]}G(x,x^{\prime},t),
\end{equation}
where the operator $\widehat{I}(t)$ has two components which are independent integrals of
motion. This system of equations together with equation \eqref{e-Gt} and the initial condition
\eqref{e-Gi} provides the complete system of equations determining the Green function. The
normalization constant for the Green function can be found using the nonlinear equation for
the evolution operator
\begin{equation}
    \widehat{U}(t) = \widehat{U}(t-\tau)\widehat{U}(\tau).
\end{equation}
This operator equality gives the nonlinear integral equation for the Green function
\begin{equation}\label{e-GG}
    G(x,x^{\prime},t) = \int G(x,y,t-\tau)G(y,x^{\prime},\tau)\,dy.
\end{equation}
This equality is equivalent to the initial condition \eqref{e-Gi} for the Green function.

The considered general properties of the quantum integrals of motion and Green function of
Schr\"{o}dinger evolution equation will be studied for partial quantum system. We consider a
particle in a uniform electric filed $F$. The Hamiltonian $\widehat{H}$ of this particle reads
\begin{equation}
    \widehat{H} = \frac{\widehat{p}^2}{2m} - F\widehat{x}.
\end{equation}
If it were the classical particle we would have the trajectory of the particle in the phase
space of the form
\begin{align}
    x &= x_0 + \frac{p_0}{m}t + \frac{Ft^2}{2m}, \\
    p &= p_0 + Ft
\end{align}
here $x$ is the coordinate and $p$ is the momentum of the classical particle, respectively.
Hence, the inverse expressions
\begin{align}
    x_0 &= x - \frac{p}{m}t + \frac{Ft^2}{2m}, \label{e-x0}\\
    p_0 &= p - Ft                              \label{e-p0}
\end{align}
provide the integrals of (classical) motion. It is easy to check that operators
\begin{align}
    \widehat{x}_0 &= \widehat{x} - \frac{\widehat{p}}{m}t + \frac{Ft^2}{2m}, \\
    \widehat{p}_0 &= \widehat{p} - Ft
\end{align}
are integrals of (quantum) motion, i.e. they satisfy the equation \eqref{e-I}. These operators
are obtained by means of quantization procedure applied to the classical integrals of motion.
The physical meaning of the discussed integrals of motion is explained by the fact that these
operators describe the initial point in the particle phase space.

For Green function $G(x,x^{\prime},t)$ we have the system of two equations
\cite{malkin-manko-dynamical-symmetries-coherent-states-quantum-systems}
\begin{align}
\begin{split}
    {\widehat{x}_0(t)}_{[x]}G(x,x^{\prime},t) &= x^{\prime}G(x,x^{\prime},t), \\
    {\widehat{p}_0(t)}_{[x]}G(x,x^{\prime},t) &= i\hbar\frac{\partial G}{\partial
    x^{\prime}}(x,x^{\prime},t),
\end{split}
\end{align}
or for the charge moving in electric field
\begin{align}\label{e-G}
\begin{split}
    \left(x + \frac{i\hbar t}{m}\frac{\partial}{\partial x} +
    \frac{Ft^2}{2m}\right)G(x,x^{\prime},t) &= x^{\prime}G(x,x^{\prime},t), \\
    \left(i\hbar\frac{\partial}{\partial x} + Ft\right)G(x,x^{\prime},t) &=
    -i\hbar\frac{\partial G}{\partial x^{\prime}}(x,x^{\prime},t)
\end{split}
\end{align}
with the initial condition \eqref{e-Gi}
\begin{equation}\label{e-G0}
    G(x,x^{\prime},0) = \delta(x-x^{\prime}).
\end{equation}
The system of equations is obtained using the quantum integrals of motion. To solve this
system we try to find the Green function in the Gaussian form, i.e.
\begin{equation}\label{e-Gexp}
    G(x,x^{\prime},t) = N(x^{\prime},t)e^{a(x^{\prime},t)x^2 + b(x^{\prime},t)x},
\end{equation}
where $N(x^{\prime},t)$, $a(x^{\prime},t)$, and $b(x^{\prime},t)$ are unknown functions.
Substituting this expression into the first equation of the system \eqref{e-G} we have the
equality
\begin{equation}
    x + \frac{i\hbar t}{m}\Bigl(2a(x^{\prime},t)x+b(x^{\prime},t)\Bigr) + \frac{Ft^2}{2m} =
    x^{\prime},
\end{equation}
from which we obtain the functions $a(x^{\prime},t)$ and $b(x^{\prime},t)$ in the form
\begin{equation}
    a(x^{\prime},t) = \frac{im}{2\hbar t}, \qquad
    b(x^{\prime},t) = -\frac{im}{\hbar t}x^{\prime} + \frac{iFt}{2\hbar}.
\end{equation}
To find out the function $N(x^{\prime},t)$ we substitute \eqref{e-Gexp} into the second
equation of the system \eqref{e-G}. As a result we obtain the equation for this function:
\begin{equation}
    \frac{\partial N}{\partial x^{\prime}} =
    \frac{i}{\hbar}\left(\frac{m}{t}x^{\prime}+\frac{1}{2}Ft\right)N.
\end{equation}
To solve this equation we try to find the function $N(x^{\prime},t)$ in the Gaussian form,
i.e.,
\begin{equation}
    N(x^{\prime},t) = n(t)e^{\alpha(t){x^{\prime}}^2+\beta(t)x^{\prime}}.
\end{equation}
Substituting this expression into previous equation we obtain the functions $\alpha(t)$ and
$\beta(t)$:
\begin{equation}
    \alpha(t) = \frac{im}{2\hbar t}, \qquad
    \beta(t) = \frac{iFt}{2\hbar}.
\end{equation}
Thus the explicit form of the Green function $G(x,x^{\prime},t)$ reads
\begin{equation}
    G(x,x^{\prime},t) = n(t)\exp\Biggl\{\frac{im}{2\hbar t}{(x-x^{\prime})}^2+
                                        \frac{iFt}{2\hbar}(x+x^{\prime})\Biggr\},
\end{equation}
where the preexponential factor is unknown function of time. To find out the function $n(t)$
we substitute this expression in the equation \eqref{e-Gt} and get the differential equation
of the form
\begin{equation}
    \frac{\partial n}{\partial t} = -\left(\frac{1}{2t} + \frac{iF^2t^2}{8m\hbar}\right)n.
\end{equation}
The solution to this equation reads
\begin{equation}
    n(t) = \frac{c}{\sqrt{t}}\exp\Biggl\{-\frac{iF^2t^3}{24m\hbar}\Biggr\}.
\end{equation}
This solution contains unknown constant $c$ and for the Green function we have the explicit
expression
\begin{equation}\label{e-Gc}
    G(x,x^{\prime},t) = \frac{c}{\sqrt{t}}
    \exp\Biggl\{\frac{im}{2\hbar t}{(x-x^{\prime})}^2+\frac{iFt}{2\hbar}(x+x^{\prime})-
    \frac{iF^2t^3}{24m\hbar}\Biggr\}.
\end{equation}
The constant $c$ can be obtained from the nonlinear equation \eqref{e-GG}. If we substitute
the expression \eqref{e-Gc} into this equation we obtain the equality
\begin{equation}
    1 = c\sqrt{\frac{2\pi i\hbar}{m}}.
\end{equation}
The above equality determines the normalization constant of the Green function
\begin{equation}
    c = \sqrt{\frac{m}{2\pi i\hbar}}.
\end{equation}
Finally we have the following expression for the Green function
\begin{equation}
    G(x,x^{\prime},t) = \sqrt{\frac{m}{2\pi i\hbar t}}
    \exp\Biggl\{\frac{im}{2\hbar t}{(x-x^{\prime})}^2+\frac{iFt}{2\hbar}(x+x^{\prime})-
    \frac{iF^2t^3}{24m\hbar}\Biggr\}.
\end{equation}
which has the Gaussian form. The expression in curle brackets is the classical action
$S(x,x^{\prime},t)$ multiplied by the factor $i/\hbar$. For $F=0$ we get the Green function
for free motion.

The equation for density operator $\widehat{\rho}(t)$ is similar to Schr\"{o}dinger eqution
for the wave function
\begin{equation}
    i\hbar\dot{\widehat{\rho}}(t) = [\widehat{H}(t),\widehat{\rho}(t)].
\end{equation}
From this equation it is easy to obtain equation for matrix elements of density matrix
$\rho(x,x^{\prime},t) = \langle x | \widehat{\rho}(t) | x^{\prime} \rangle$
\begin{equation}\label{e-rho}
    \frac{\partial\rho}{\partial t}(x,x^{\prime},t) +
    \frac{i}{\hbar}\Bigl({\widehat{H}}_{[x]}\rho(x,x^{\prime},t)-
                         {\widehat{H}^t}_{[x^{\prime}]}\rho(x,x^{\prime},t)\Bigr)
                         =0.
\end{equation}
The Hamiltonian of moving particle with potential $V(x)$ reads
\begin{equation}
    \widehat{H} = \frac{\widehat{p}^2}{2m} + V(x),
\end{equation}
and equation \eqref{e-rho} takes the form
\begin{equation}\label{e-vN}
    \frac{\partial\rho}{\partial t}(x,x^{\prime},t) -
    \Biggl(\frac{i\hbar}{2m}\left(\frac{\partial^2}{\partial x^2}-
           \frac{\partial^2}{\partial {x^{\prime}}^2}\right) -
    \frac{i}{\hbar}\Bigl(V(x)-V(x^{\prime})\Bigr)\Biggr)\rho(x,x^{\prime},t) = 0.
\end{equation}
The equation for density matrix admits the solutions in factorised form
$\psi(x,t)\psi^*(x^{\prime},t)$ which describe the pure states of the quantum system.

\begin{figure}
\begin{center}
    \includegraphics[width=75mm, height=75mm]{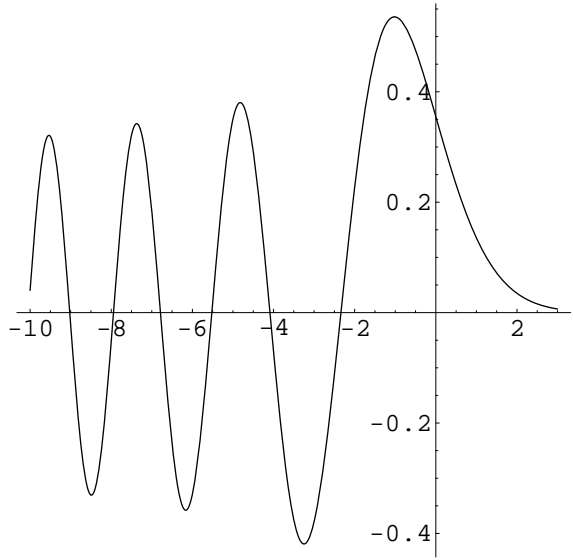}
    \caption{Airy function}\label{f-Airy}
\end{center}
\end{figure}

The Schr\"{o}dinger equation for stationary state of the particle in the electric field reads
\begin{equation}\label{e-Sch}
    \frac{d^2\psi}{dx^2} + \frac{2m}{\hbar^2}\left(\mathcal{E} + Fx\right)\psi = 0.
\end{equation}
It is well known that for all energy values $\mathcal{E}$ there exists the solution
$\psi_\mathcal{E}(x)$ of equation \eqref{e-Sch} given by a formula
\begin{equation}\label{e-psi}
    \psi_\mathcal{E}(x) = \frac{{\left(2m\right)}^{1/3}}{\pi^{1/2}F^{1/6}\hbar^{2/3}}
    \Phi\left(-{\left(\frac{2mF}{\hbar^2}\right)}^{1/3}\left(x+\frac{\mathcal{E}}{F}\right)\right),
\end{equation}
where $\Phi(x)$ is the Airy function:
\begin{equation}\label{e-Phi}
    \Phi(x) = \frac{1}{\sqrt{\pi}}\int\limits^{+\infty}_0
    \cos\left(\frac{u^3}{3} + ux\right)\,du =
    \frac{1}{2\sqrt{\pi}}\int\exp\left\{i\left(\frac{u^3}{3} +
    ux\right)\right\}\,du,
\end{equation}
and the coefficient in front of the function $\Phi$ is necessary for normalization of
$\psi_\mathcal{E}$ to $\delta$-function:
\begin{equation}
    \int\psi_\mathcal{E}(x)\psi_\mathcal{E^\prime}(x)\,dx =
    \delta\left(\mathcal{E^\prime} - \mathcal{E}\right).
\end{equation}
Plot of the Airy function is presented in Figure \eqref{f-Airy}.
Let us introduce notations
\begin{equation}
    A = \frac{{\left(2m\right)}^{1/3}}{\pi^{1/2}F^{1/6}\hbar^{2/3}},\quad
    \alpha = -\left(\frac{2mF}{\hbar^2}\right)^{1/3},\quad
    \varepsilon = \alpha\frac{\mathcal{E}}{F}.
\end{equation}
In this notations $\psi_\mathcal{E}$ reads $\psi_\mathcal{E}(x) = A\Phi\left(\alpha x +
\varepsilon\right)$.

\section{Weyl representation}
The density operator of quantum state can be considered in different reprresentations. One of
the representations is Weyl representation. In this representation each operator is described
by its symbol. The Weyl symbol of the density operator is called Wigner function.

Now we obtain the equation for the Wigner function. The Wigner function $W(q,p)$ is expressed
in terms of matrix elements of density operator as \cite{pr-40-749}
\begin{equation}
    W(q,p) = \int \rho\left(q+\frac{u}{2},q-\frac{u}{2}\right)e^{-\frac{i}{\hbar}pu}\,du.
\end{equation}
To obtain the equation for the Wigner function we have to know the correspondence between
operators acting on the functions of variables $x$, $x^{\prime}$ and the same operators acting
on the functions of variables $q$, $p$. The operators $\widehat{f}_{[x,x^{\prime}]}$ and
$\widehat{F}_{[q,p]}$ are called corresponding to each other
\begin{equation}
    \widehat{f}_{[x,x^{\prime}]}
    \longleftrightarrow
    \widehat{F}_{[q,p]}
\end{equation}
if the following equivalent relations are valid for any corresponding to each other density
matrix $\rho(x,x^{\prime})$ and the Wigner function $W(q,p)$ of the same quantum state
\begin{equation}\label{e-rhoW}
\begin{split}
   &W(q,p) = \int \rho\left(q+\frac{u}{2},q-\frac{u}{2}\right)e^{-\frac{i}{\hbar}pu}\,du
    \Rightarrow \\
   &\int\left(\widehat{f}_{[x,x^{\prime}]}\rho\right)
    \left(q+\frac{u}{2},q-\frac{u}{2}\right)e^{-\frac{i}{\hbar}pu}\,du =
    \left(\widehat{F}_{[q,p]}W\right)(q,p),
\end{split}
\end{equation}
and
\begin{equation}\label{e-Wrho}
\begin{split}
   &\rho(x,x^{\prime}) = \frac{1}{2\pi\hbar}\int W\left(\frac{x+x^{\prime}}{2},p\right)
    e^{\frac{i}{\hbar}p(x-x^{\prime})}\,dp
    \Rightarrow \\
   &\left(\widehat{f}_{[x,x^{\prime}]}
    \rho\right)(x,x^{\prime}) = \frac{1}{\hbar}
    \int\left(\widehat{F}_{[q,p]}W\right)
    \left(\frac{x+x^{\prime}}{2},p\right)e^{\frac{i}{\hbar}p(x-x^{\prime})}\,dp.
\end{split}
\end{equation}
Now we obtain operators corresponding to the operators $\widehat{x}$, $\widehat{x^{\prime}}$,
$\partial/\partial x$, and $\partial/\partial x^{\prime}$.
\begin{equation}
\begin{split}
    \int\left(\widehat{x}\rho\right)\left(q+\frac{u}{2},q-\frac{u}{2}\right)
    e^{-\frac{i}{\hbar}pu}\,du &=
    \int\left(q+\frac{u}{2}\right)\rho\left(q+\frac{u}{2},q-\frac{u}{2}\right)
    e^{-\frac{i}{\hbar}pu}\,du \\
    &= \biggl(q+\frac{1}{2}\left(-\frac{\hbar}{i}\frac{\partial}{\partial
    p}\right)\biggr)W(q,p).
\end{split}
\end{equation}
It means that multiplication of density matrix by the coordinate $x$ provides the action on
the Wigner function of the form
\begin{equation}
    x\rho(x,x^{\prime}) \longleftrightarrow
    \left(q+\frac{i\hbar}{2}\frac{\partial}{\partial p}\right)W(q,p).
\end{equation}
Analogously one can obtain that
\begin{equation}
    x^{\prime}\rho(x,x^{\prime}) \longleftrightarrow
    \left(q-\frac{i\hbar}{2}\frac{\partial}{\partial p}\right)W(q,p).
\end{equation}
To obtain operators acting on the Wigner function corresponding to operators $\widehat{x}$ and
$\widehat{x^{\prime}}$ we used the equation \eqref{e-rhoW}. Now to obtain operators acting on
the Wigner function corresponding to operators $\partial/\partial x$ and $\partial/\partial
x^{\prime}$ acting on the density matrix we use the equation \eqref{e-Wrho} and get
\begin{equation}
    \frac{\partial\rho}{\partial x}(x,x^{\prime}) =
    \frac{1}{2\pi\hbar}\int\left(\frac{1}{2}\frac{\partial}{\partial q} + \frac{i}{\hbar}p\right)
    W\left(\frac{x+x^{\prime}}{2},p\right)e^{\frac{i}{\hbar}p(x-x^{\prime})}\,dp.
\end{equation}
By definition of corresponding operators we have
\begin{equation}
    \frac{\partial}{\partial x}\rho(x,x^{\prime}) \longleftrightarrow
    \left(\frac{1}{2}\frac{\partial}{\partial q} + \frac{i}{\hbar}p\right)W(q,p),
\end{equation}
and analogously
\begin{equation}
    \frac{\partial}{\partial x^{\prime}}\rho(x,x^{\prime}) \longleftrightarrow
    \left(\frac{1}{2}\frac{\partial}{\partial q} - \frac{i}{\hbar}p\right)W(q,p).
\end{equation}
Now using the obtained correspondence rules and the von Neumenn equation \eqref{e-vN} for the
density matrix we can write the equation for the Wigner function, which was obtained by Moyal
\cite{pcps-45-99}
\begin{equation}
    \frac{\partial W}{\partial t}(q,p,t) +\frac{p}{m}\frac{\partial W}{\partial q}(q,p,t) +
    \frac{i}{\hbar}\Biggl(V\left(q+\frac{i\hbar}{2}\frac{\partial}{\partial p}\right) -
                          V\left(q-\frac{i\hbar}{2}\frac{\partial}{\partial p}\right)\Biggr)
    W(q,p,t) = 0.
\end{equation}
In the case of charged particle moving in electric field the potential energy $V(x) = -Fx$,
and one gets the Moyal equation in the form
\begin{equation}
    \dot{W} + \frac{p}{m}\frac{\partial W}{\partial q} + \frac{1}{2}F\frac{\partial W}{\partial p} =
    0.
\end{equation}

The Wigner function $W(q,p,t)$ can be found as solution to the Moyal equation. It has the form
\begin{equation}
    W(q,p,t) = W_0(x_0,p_0)
\end{equation}
where $W_0(q,p)$ is the initial Wigner function and $x_0$ and $p_0$ are given by equations
\eqref{e-x0} and \eqref{e-p0}, respectively. We get
\begin{equation}
    W(q,p,t) = W_0\left(q-\frac{p}{m}t+\frac{Ft^2}{2m},p-Ft\right).
\end{equation}

For Wigner function one can introduce the propagator $\Gamma(q,p,q^{\prime},p^{\prime},t)$ of
Moyal equation
\begin{equation}
    W(q,p,t) = \int
    \Gamma(q,p,q^{\prime},p^{\prime},t)W(q^{\prime},p^{\prime},0)\,dq^{\prime}\,dp^{\prime}.
\end{equation}
In the case of the charge moving in electric field we have explicit expression for the
propagator of Moyal equation
\begin{equation}
    \Gamma(q,p,q^{\prime},p^{\prime},t) = \delta(x_0-q^{\prime})
    \delta(p_0-p^{\prime}) = \delta\left(q-\frac{p}{m}t-q^{\prime} +\frac{Ft^2}{2m}\right)
    \delta\left(p-p^{\prime}-Ft\right).
\end{equation}
Let us consider example of initial state which is described by the wave function
\begin{equation}
    \psi_0(x) =
    \sqrt[4]{\frac{m\omega}{\pi\hbar}}\exp\Biggl\{-\frac{m\omega}{2\hbar}x^2\Biggr\}.
\end{equation}
This state can be specially prepared by means of a spring which is removed at the time moment
$t=0$. The corresponding Wigner function has the initial form
\begin{equation}
    W_0(q,p) = 2\exp\left\{-\frac{m\omega}{\hbar}q^2-\frac{p^2}{\hbar m\omega}\right\}.
\end{equation}
The evolution of the Wigner function is given by formula
\begin{equation}
\begin{split}
    W(q,p,t) &= 2\exp\Biggl\{-\frac{m\omega}{\hbar}q^2+\frac{2\omega t}{\hbar}qp-
    \frac{1}{\hbar m}\left(\omega t^2+\frac{1}{\omega}\right)p^2-\frac{Ft^3\omega}{\hbar m}p\Biggr.\\
    \Biggl.&-\frac{Ft^2\omega}{\hbar}q-\frac{F^2t^2}{\hbar m}\left(\frac{\omega t^2}{4}+\frac{1}{\omega}\right)+
    \frac{2Ft}{\hbar m\omega}\Biggr\}.
\end{split}
\end{equation}

\begin{figure}
\begin{center}
    \includegraphics{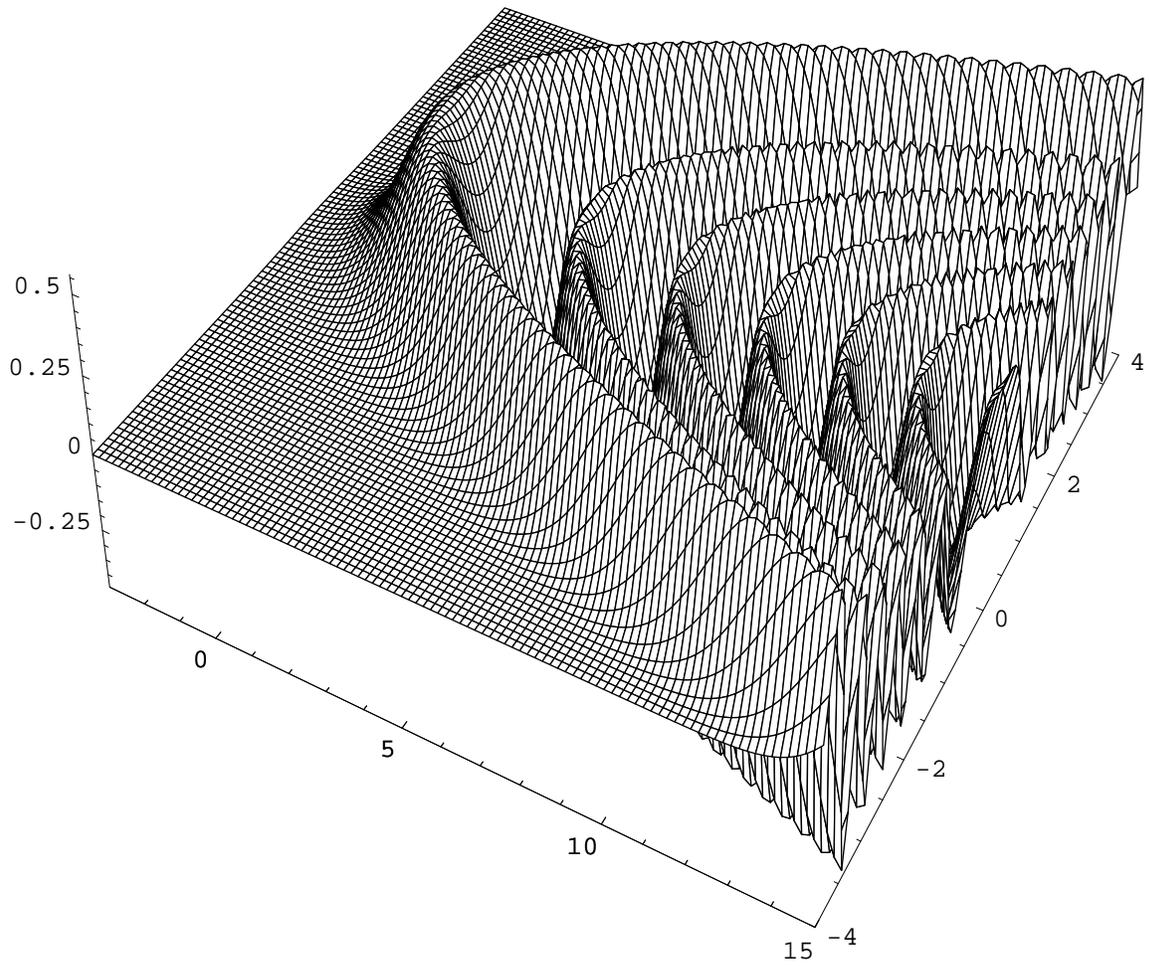}
    \caption{Wigner function}\label{f-Wig}
\end{center}
\end{figure}

Now we calculate the Wigner's function of the stationary state of the charged particle with
given energy \eqref{e-psi}. By definition, Wigner function of the state is related to the wave
function $\psi_\mathcal{E}(x)$ by the formula
\begin{equation}
    W_\mathcal{E}(q,p) = \int\psi_\mathcal{E}\left(q+\frac{u}{2}\right)
                             \psi^*_\mathcal{E}\left(q-\frac{u}{2}\right)e^{-\frac{i}{\hbar}pu}\,du.
\end{equation}
Using expressions \eqref{e-psi} and \eqref{e-Phi} it is easy to show that
\begin{equation}
\begin{split}
    W_\mathcal{E}(q,p) &= \frac{1}{\pi}\sqrt[3]{\frac{m}{F^2\hbar^2}}\Phi\left(\sqrt[3]{4}\left(\frac{p^2}{\alpha^2\hbar^2}
                         + \alpha q + \varepsilon\right)\right) \\
                       &= \frac{1}{\pi}\sqrt[3]{\frac{m}{F^2\hbar^2}}\Phi\left({
                         \left(\frac{1}{mF\hbar}\right)}^{2/3}p^2 - 2{\left(\frac{mF}{\hbar^2}\right)}^{1/3}q
                         -2{\left(\frac{m}{F^2\hbar^2}\right)}^{1/3}\mathcal{E}\right).
\end{split}
\end{equation}
Plot of the Wigner function is shown in Figure \eqref{f-Wig}.

\section{Tomographic representation}

The tomogram of the quantum state is expressed in terms of Wigner function ($\hbar=1$)
\cite{qso-7-615}.
\begin{equation}
    w(X,\mu,\nu) = \int W(q,p)e^{ik(X-\mu q-\nu p)}\,\frac{dk\,dq\,dp}{{(2\pi)}^2}.
\end{equation}
The inverse relation determines the Wigner function in terms of the tomogram
\begin{equation}
    W(q,p) = \int w(X,\mu,\nu) e^{i(X-\mu q-\nu p)}\,\frac{dX\,d\mu\,d\nu}{2\pi}.
\end{equation}
For pure state described by wave function $\psi(x)$ for tomogram one can obtain the formula in
terms of this wave function
\begin{equation}\label{e-wpsi}
    w(X,\mu,\nu) = \frac{1}{2\pi|\nu|}{\left|\int\psi(y)\exp\left\{\frac{i\mu}{2\nu}y^2-
    \frac{iX}{\nu}y\right\}\,dy\right|}^2.
\end{equation}

To write the equation for tomogram \cite{pl-a-213-1} we have to know the correspondence
between operators acting on Wigner functions depending on the variables $q$, $p$ and the same
operators acting on the tomograms depending on the variables $X$, $\mu$, $\nu$. Operators
$\widehat{F}_{[q,p]}$ and $\widehat{G}_{[X,\mu, \nu]}$ are called corresponding to each other
\begin{equation}
    \widehat{F}_{[q,p]}
    \longleftrightarrow
    \widehat{G}_{[X,\mu,\nu]}
\end{equation}
if the following equivalent relations are valid for any corresponding to each other Wigner
function $W(q,p)$ and tomogram $w(X,\mu,\nu)$
\begin{equation}
\begin{split}
   &w(X,\mu,\nu) = \int W(q,p)e^{-ik(X-\mu q-\nu p)}\,\frac{dk\,dq\,dp}{{(2\pi)}^2} \Rightarrow \\
   &\int \left(\widehat{F}_{[q,p]}W\right)(q,p)
    e^{-ik(X-\mu q-\nu p)}\,\frac{dk\,dq\,dp}{{(2\pi)}^2} =
    \left(\widehat{G}_{[X,\mu,\nu]}w\right)(X,\mu,\nu)
\end{split}
\end{equation}
and
\begin{equation}
\begin{split}
   &W(q,p) = \int w(X,\mu,\nu)e^{i(X-\mu q-\nu p)}\,\frac{dX\,d\mu\,d\nu}{2\pi} \Rightarrow \\
   &\left(\widehat{F}_{[q,p]}W\right)(q,p) =
    \int \left(\widehat{G}_{[X,\mu,\nu]}w\right)(X,\mu,\nu)
    e^{i(X-\mu q-\nu p)}\,\frac{dX\,d\mu\,d\nu}{2\pi}.
\end{split}
\end{equation}
Using the same method which we used for finding the corresponding operators in position and
Weyl representation we can get the following relations
\begin{alignat}{2}
    qW(q,p) &\longleftrightarrow -{\left(\frac{\partial}{\partial
    X}\right)}^{-1}\frac{\partial}{\partial\mu}w(X,\mu,\nu), &\qquad
    \frac{\partial}{\partial q}W(q,p) &\longleftrightarrow \mu\frac{\partial}{\partial X}w(X,\mu,\nu), \\
    pW(q,p) &\longleftrightarrow -{\left(\frac{\partial}{\partial
    X}\right)}^{-1}\frac{\partial}{\partial\nu}w(X,\mu,\nu), &\qquad
    \frac{\partial}{\partial p}W(q,p) &\longleftrightarrow \nu\frac{\partial}{\partial X}w(X,\mu,\nu).
\end{alignat}
The action of the operators onto homogeneous tomograms must give the homogeneous function.
This condition gives the possibility to avoid the ambiguity related with the fact that Wigner
function depends on two variables and the tomogram depends on the three variables.

The evolution equation for tomogram in which we reconstruct the Plank constant reads
\cite{pl-a-213-1}
\begin{equation}
    \frac{\partial w}{\partial t} - \mu\frac{\partial w}{\partial\nu} +
    \frac{i}{\hbar}\Biggl[V\left(-{\left(\frac{\partial}{\partial X}\right)}^{-1}\frac{\partial}{\partial\mu}+
    \frac{i\hbar\nu}{2}\frac{\partial}{\partial X}\right) -
    V\left(-{\left(\frac{\partial}{\partial X}\right)}^{-1}\frac{\partial}{\partial\mu}-
    \frac{i\hbar\nu}{2}\frac{\partial}{\partial X}\right)\Biggr]w = 0.
\end{equation}
In our case of charged particle moving in electric field the equation takes the simple form
\begin{equation}
    \frac{\partial w}{\partial t} - \mu\frac{\partial w}{\partial\nu} +
    F\nu\frac{\partial}{\partial X}w = 0.
\end{equation}

Now we find thetomographic symbols of the integrals of motion. The kernels of the integrals of
motion in position representation read
\begin{equation}
\begin{split}
    x_0(x,x^{\prime}) &= \left(x+\frac{Ft^2}{2m}\right)\delta(x-x^{\prime}) -
    \frac{i\hbar t}{m}\delta^{\prime}(x-x^{\prime}), \\
    p_0(x,x^{\prime}) &= -Ft\delta(x-x^{\prime})+i\hbar\delta^{\prime}(x-x^{\prime}).
\end{split}
\end{equation}
Using these kernels one can find Weyl symbols of the integrals of motion
\begin{equation}
\begin{split}
    W_{\widehat{x}_0}(q,p) &= \int x_0\left(q+\frac{u}{2},q-\frac{u}{2}\right)
    e^{-\frac{i}{\hbar}pu}\,du = q - \frac{p}{m}t + \frac{Ft^2}{2m}, \\
    W_{\widehat{p}_0}(q,p) &= \int p_0\left(q+\frac{u}{2},q-\frac{u}{2}\right)
    e^{-\frac{i}{\hbar}pu}\,du = p -Ft.
\end{split}
\end{equation}
The tomographic symbols of the integrals of motion are related the their Weyl symbols by the
same formulas that the tomogram of a quantum state is related to the Wigner function
($\hbar=1$), i.e.,
\begin{equation}
\begin{split}
    w_{\widehat{x}_0}(X,\mu,\nu) &= \int W_{\widehat{x}_0}(q,p)
    e^{-ik(X-\mu q-\nu p)}\,\frac{dk\,dq\,dp}{{(2\pi)}^2} \\
    &= \Biggl(-{\left(\frac{\partial}{\partial X}\right)}^{-1}\frac{\partial}{\partial\mu} -
    \frac{t}{m}{\left(\frac{\partial}{\partial X}\right)}^{-1}\frac{\partial}{\partial\nu} +
    \frac{Ft^2}{2m}\Biggr)w_{\widehat{1}}(X,\mu,\nu), \\
    w_{\widehat{p}_0}(X,\mu,\nu) &= \int W_{\widehat{p}_0}(q,p)
    e^{-ik(X-\mu q-\nu p)}\,\frac{dk\,dq\,dp}{{(2\pi)}^2} \\
    &= -\Biggl({\left(\frac{\partial}{\partial X}\right)}^{-1}\frac{\partial}{\partial\nu} +
    Ft\Biggr)w_{\widehat{1}}(X,\mu,\nu).
\end{split}
\end{equation}
The tomographic symbol of identity operator in these equalities has the form
\begin{equation}
    w_{\widehat{1}}(X,\mu,\nu) = \int\frac{e^{-ikX}}{k^2}\,dk\,\delta(\mu)\delta(\nu)
\end{equation}

The evolution equation for the tomogram can be solved by means of the Green function
(propagator) of this equation relating the initial and current tomograms by the integral
transform
\begin{equation}
w(X,\mu,\nu,t) = \int \Pi(X,\mu,\nu,X^{\prime},\mu^{\prime},\nu^{\prime},t)
                 w(X^{\prime},\mu^{\prime},\nu^{\prime},0)\,dX^{\prime}\,d\mu^{\prime}\,d\nu^{\prime}.
\end{equation}
The propagator satisfies the evolution equation (we reconstruct the Plank constant)
\begin{equation}
    \frac{\partial \Pi}{\partial t} - \mu\frac{\partial \Pi}{\partial\nu} +
    \frac{i}{\hbar}\Biggl[V\left(-{\left(\frac{\partial}{\partial X}\right)}^{-1}\frac{\partial}{\partial\mu}+
    \frac{i\hbar\nu}{2}\frac{\partial}{\partial X}\right) -
    V\left(-{\left(\frac{\partial}{\partial X}\right)}^{-1}\frac{\partial}{\partial\mu}-
    \frac{i\hbar\nu}{2}\frac{\partial}{\partial X}\right)\Biggr]\Pi = 0,
\end{equation}
and the initial condition
\begin{equation}
    \Pi(X,\mu,\nu,X^{\prime},\mu^{\prime},\nu^{\prime},0) =
    \delta(X-X^{\prime})\delta(\mu-\mu^{\prime})\delta(\nu-\nu^{\prime}).
\end{equation}
It can be shown \cite{jrlr-18-407, jrlr-17-579} that there is the relation of the Green
functions of Schr\"{o}dinger and tomographic evolution equations ($\hbar=1$)
\begin{equation}
\begin{split}
    \Pi(X,\mu,\nu,X^{\prime},\mu^{\prime},\nu^{\prime},t) &=
    \frac{1}{4\pi^2}\int k^2 G\left(a+\frac{k\nu}{2},y,t\right)
    G^*\left(a-\frac{k\nu}{2},z,t\right)\delta(y-z-k\nu^{\prime})\\
    &\times\exp\biggl\{ik\left(X^{\prime}-X+\mu a -
    \mu^{\prime}\frac{y+z}{2}\right)\biggr\}\,dk\,dy\,dz\,da.
\end{split}
\end{equation}
For the charge moving in the electric field the evolution equation for the propagator reads
\begin{equation}
    \frac{\partial \Pi}{\partial t} - \mu\frac{\partial \Pi}{\partial\nu} +
    F\nu\frac{\partial\Pi}{\partial X} = 0.
\end{equation}
It has solution
\begin{equation}
    \Pi(X,\mu,\nu,X^{\prime},\mu^{\prime},\nu^{\prime},t) = \delta(X-X^{\prime}-Ft)
    \delta\left(\mu-\mu^{\prime}-\frac{t}{m}\nu\right)\delta(\nu-\nu^{\prime}).
\end{equation}

\begin{figure}
\begin{center}
    \includegraphics{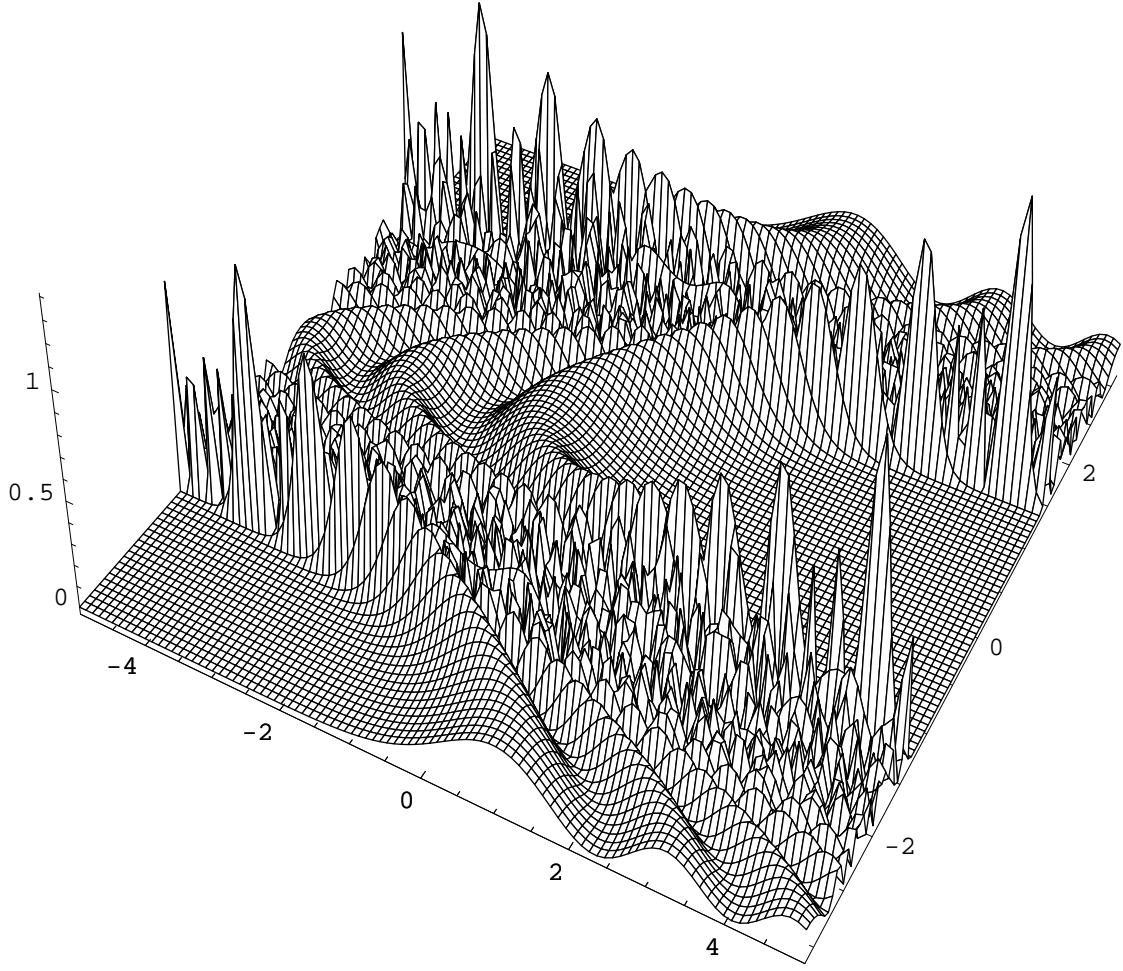}
    \caption{Tomogram}\label{f-Tom}
\end{center}
\end{figure}

It is interesting to calculate the tomographic probability distribution of stationary state of
the charged particle.  Evaluating the corresponding integral \eqref{e-wpsi} we get the
tomogram of the state
\begin{equation}
    w_\mathcal{E}\left(X,\mu,\nu\right) = \frac{1}{2\pi|\nu|}{\left|\int\psi_\mathcal{E}(y)
                                         \exp\left\{\frac{i\mu}{2\hbar\nu}y^2-\frac{iX}{\hbar\nu}y\right\}\,dy\right|}^2.
\end{equation}
Using explicit expressions for $\psi_\mathcal{E}$ one can obtain the explicit expression for
the tomogram
\begin{equation}
\begin{split}
    w_\mathcal{E}\left(X,\mu,\nu\right) &= \frac{A^2\hbar}{4\pi|\mu|}{\left|\Phi\left(\varepsilon-\frac{\alpha X}{\mu}-
                                          \frac{\hbar^2\alpha^2}{4}\frac{\nu^2}{\mu^2}\right)\right|}^2 \\
                                        &= \sqrt[3]{\frac{m^2}{16\pi^2F\hbar}}\frac{1}{|\mu|}
                                        {\left|\Phi\left( -{\left(\frac{2m}{F^2\hbar^2}\right)}^{1/3}\mathcal{E}
                                        +{\left(\frac{2mF}{\hbar^2}\right)}^{1/3}\frac{X}{\mu}
                                        -{\left(\frac{m^2F^2}{2\hbar}\right)}^{2/3}\frac{\nu^2}{\mu^2}\right)\right|}^2.
\end{split}
\end{equation}
Plot of the tomogram is shown in Figure \eqref{f-Tom}. One can see the reach structure of the
tomogram.

\section{Conclusion}

Tha main result of our work is the obtaining the tomogram of the stationary state of the
charged particle moving in the constant electric field. We found closed connection with
integrals of motion, Wigner function and tomograms. The obtained results give a possibility to
find new relations for the Airy function.

\section{Acknowledgements}

This work was supported by the Russian Foundation for Basic Research under Projects Nos.
$99$--$02$--$17753$ and $01$--$02$--$17745$. E. V. S thanks Professor T. Seligman for
providing a fellowship.

\end{document}